\documentclass[12pt]{article}
\usepackage{color}
\usepackage{amsmath, amssymb}
\usepackage{epsfig, palatino}
\usepackage{pstricks,pst-node,pst-tree}
\usepackage{epic}

\usepackage{mathrsfs}

\usepackage{ae} 
\usepackage[T1]{fontenc}
\usepackage[ansinew]{inputenc}
\usepackage{amsmath}
\usepackage{amssymb}
\usepackage{graphicx}
\usepackage{color}
\definecolor{darkblue}{cmyk}{0.9,0.9,0,0}
\usepackage[colorlinks=true,linkcolor=darkblue,citecolor=darkblue,urlcolor=darkblue]{hyperref}

\newcommand{\beq}{\begin{equation}}
\newcommand{\eeq}{\end{equation}}
\newcommand\beqa{\begin{eqnarray}}
\newcommand\eeqa{\end{eqnarray}}
\newcommand\bea{\begin{array}}
\newcommand\eea{\end{array}}

\def\XXint#1#2#3{{\setbox0=\hbox{$#1{#2#3}{\int}$}
\vcenter{\hbox{$#2#3$}}\kern-.5\wd0}}

\newcommand{\nn}{\nonumber}

\newcommand{\COMMENT}[1]{}

\newcommand{\neqa}{\nonumber\end{eqnarray}}
\newcommand{\la}[1]{\label{#1}}

\renewcommand{\d}{\partial}

\newcommand{\<}{{\langle}}
\renewcommand{\>}{{\rangle}}

\newcommand{\re}{\relax{\rm I\kern-.18em R}}

\def\su2{{SU(2)}}

\def\[{\left[}
\def\]{\right]}

\def\({\left(}
\def\){\right)}
\def\[{\left[}
\def\]{\right]}

\def\<{\langle}
\def\>{\rangle}

\def\i2{\frac{i}{2}}

\def\cF{{\cal F}}
\def\cC{{\cal C}}
\def\cR{{\cal R}}

\usepackage{varioref}
\usepackage{makeidx}
\makeindex
\usepackage[english]{babel}

\begin{document}

%

\thispagestyle{empty}

\renewcommand{\thefootnote}{\fnsymbol{footnote}}
\setcounter{footnote}{0}
\setcounter{figure}{0}
\begin{center}
$$$$
{\Large\textbf{\mathversion{bold}
OPE for Super  Loops
}\par}

\vspace{1.0cm}
\textrm{Amit Sever{$^\tau$}, Pedro Vieira{$^\sigma$}, Tianheng Wang{$^\phi$}}
\\ \vspace{1.2cm}
\footnotesize{
\textit{
Perimeter Institute for Theoretical Physics\\ Waterloo,
Ontario N2J 2W9, Canada} \\
\texttt{}
\vspace{3mm}
}

$\{\tau,\sigma,\phi\}$\verb" = {amit.sever@gmail.com, pedrogvieira@gmail.com, twang@perimeterinstitute.ca}"
\par\vspace{2cm}

\textbf{Abstract}
\end{center}

We extend the Operator Product Expansion for Null Polygon Wilson loops to the Mason-Skinner-Caron-Huot super loop dual to non MHV gluon amplitudes.  We explain how the known tree level amplitudes can be promoted into an infinite amount of data at any loop order in the OPE picture. As an application, we re-derive all one loop NMHV six gluon amplitudes by promoting their tree level expressions. We also present some new all loops predictions for these amplitudes.    

\vspace*{\fill}

\setcounter{page}{1}
\renewcommand{\thefootnote}{\arabic{footnote}}
\setcounter{footnote}{0}

\newpage

\tableofcontents
\section{Introduction}

Four dimensional gauge theories are, secretly,  string theories \cite{'tHooft:1973jz}. This is really exciting both conceptually as well as pragmatically. In practice it means that if one is smart enough one can map four dimensional computations into two dimensional problems; these are of course way simpler. A remarkable property of some gauge theories is that they are integrable meaning that their two dimensional 't Hooft worldsheet is exactly solvable. The most known example is $\mathcal{N}=4$ SYM also popularly known as the \textit{Harmonic Oscillator of Gauge Theories} \cite{BPRetc}. An object which has been under attack from several fronts is the planar S-matrix  of this theory. 
One approach towards addressing the computation of scattering amplitudes using integrability, known as the Operator Product Expansion (OPE) \cite{OPEpaper}, goes as follows 

\begin{figure}[t]
\centering
\def\svgwidth{10cm}
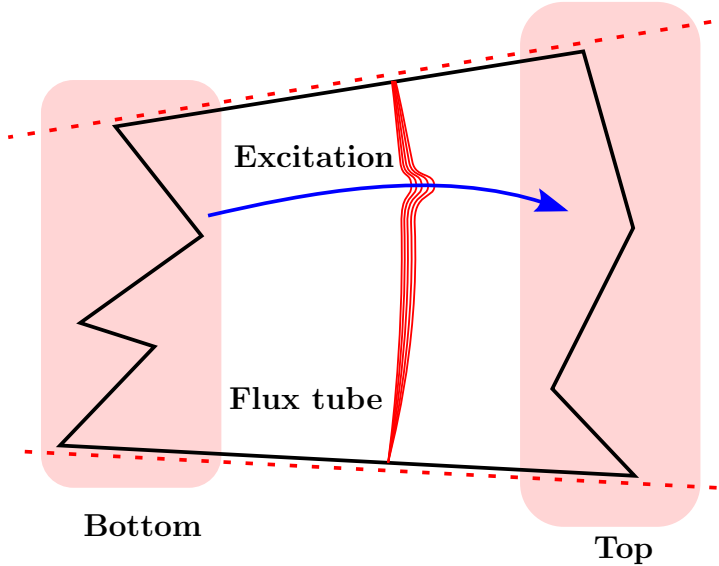
\caption{\small The OPE picture. \textit{Flux tube  excitations} are created at the \textit{bottom} and absorbed in the \textit{top}. The flux is tick in spacetime but is 1+1 dimensional in AdS. Its excitations are integrable.}\label{OPEpicture}
\end{figure}

\begin{figure}[t]
\centering
\def\svgwidth{12cm}
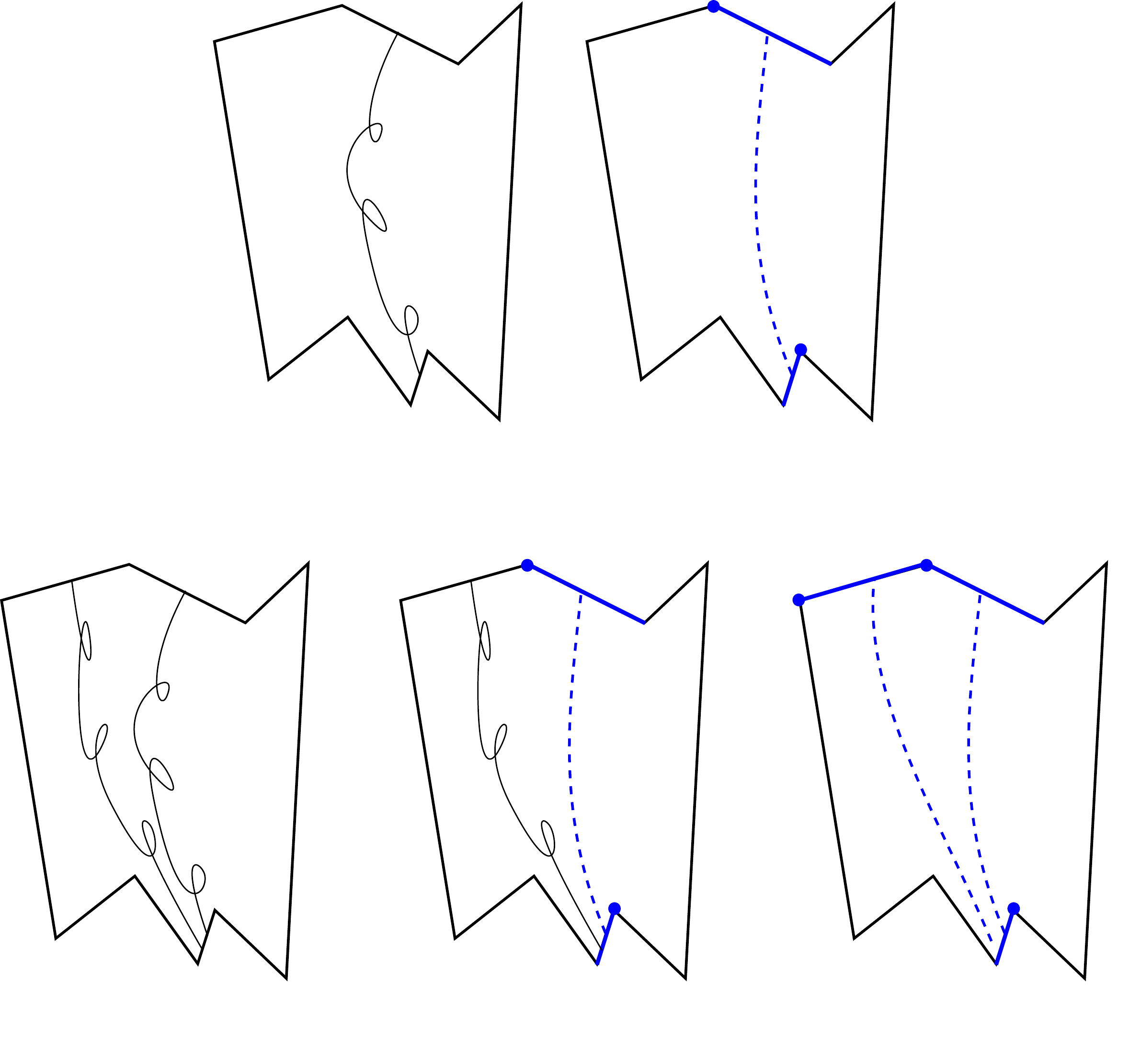
\caption{\small One remarkable property of the Super Wilson Loop/Scattering Amplitude duality is that a N$^k$MHV scattering amplitude at $l$ loops is given, roughly, by a $k+l$ Wilson loop computation. This means that the known tree and one loop data from Scattering amplitude present an excellent window towards higher loop physics from the Wilson loop point of view. In particular, multi-particle flux tube excitations can in principle be studies with the already available data.}\label{examples}
\end{figure}

\begin{itemize}
\item First we make use of a remarkable duality which tells us that planar scattering amplitudes are computed by the expectation value of (some supersymmetric) Null Polygon Wilson loops (NPWL) \cite{AmplitudeWilson,Skinner,Simon,super2}. Maximally Helicity Violating (MHV) amplitudes are computed by the usual bosonic Wilson loop \cite{AmplitudeWilson}. Amplitudes with other helicities are given by a supersymmetric version of this loop \cite{Skinner,Simon}.
\item Next, and most importantly, one needs to rephrase the problem of computing the four dimensional amplitude into a manifestly two dimensional language. To do so, we single out two of the edges of the polygon loop. We think of them as a quark/anti-quark pair moving at the speed of light and sourcing a color flux tube stretching between them. This color flux is thick in spacetime. However, it has an holographic description in terms of a two dimensional object \cite{AMcomments}. That is the two dimensional description we were looking for.
\item The choice of the two edges splits the polygon into a top and a bottom parts. We  interpret the computation of the NPWL as describing the creation of excitations in the bottom part of the polygon which propagate in the flux tube and then absorbed by the top part of polygon, see figure \ref{OPEpicture}. In other words we compute a transition amplitude in the two dimensional system.
\item The computation is now taking place in one (plus one) dimension. The energy of the flux tube excitations in planar $\mathcal{N}=4$ SYM can be studied to all loop orders using integrability \cite{Benjamin,otherpapers,Benjamin2}. 
This means that the propagation part of the computation is under good control. The computation of the creation/absorption form factors within the integrability formalism is not yet known. A strategy to overcome this difficulty is to predict a big part of the result,  coming from corrections to the propagation part in all possible channels.\footnote{More precisely, in the OPE approach, we compute the discontinuity of the Wilson loop in some conformal cross-ratios $u_i$. Similarly, the Unitarity method \cite{Zvi} for computing scattering amplitudes is based on the computation of discontinuities in Mandelstam invariants. It would be interesting to find a concrete relation between the two. This was partially achieved in the very interesting recent work \cite{SimonNew} where discontinuities of Feynman diagrams w.r.t. cross-ratios were taken.} That is, for all possible choices of pairs of edges. Then we try to bootstrap the full result  by imposing crossing symmetry. 
\end{itemize}
This program was carried out in \cite{bootstraping,Hexagonpaper,heptagon} for MHV amplitudes 
, that is for bosonic Wilson loops.
Using this method, two loop results \cite{DelDucaOct,DelDuca,Gon} obtained from direct computations were reproduced and infinitely many higher loop predictions were presented. One feature of all these OPE computations is that only single particle perturbations of the flux tube were considered.\footnote{This resembles a lot the leading log expansion in Reggee theory and BFKL \cite{BFKL}. It would be very interesting to establish a more solid connection, see \cite{Bartels:2011xy} for interesting advances in this direction.} To go beyond we need to better understand how multi-particle flux excitations behave. In principle, with integrability, all we need to know is how two particles interact. The scattering of $N$ particles then factorizes in a sequence of $(N-1)!$ two body scattering events. From the MHV point of view, the critical missing point of data which, optimistically, should shed light over the full integrability structure is a three loop scattering amplitude.\footnote{The simplest of such amplitudes is probably the three loop eight point amplitude in a restricted kinematics regime where all momenta of the external particles lie in the same plane \cite{AMapril}. } 
That  seems like a very hardcore computation with the current Feynman diagramatic tools. 

In this paper we initiate the study of non MHV amplitudes from the OPE point of view. Our main motivation resides on a remarkable feature of the Wilson Loop/Scattering amplitude duality which is that a N$^k$MHV scattering amplitude at  $l$ loops is given by a Wilson loop computation at, roughly, $l+k$ loops \cite{Skinner,Simon}.\footnote{The fact that constant complexity slices are roughly the $l+k = constant$ slices is also a feature of the all loop recursion relations \cite{Integrand}. In fact, it would be very interesting to understand how the OPE expansion hides inside the integrand. From the OPE point of view the Amplitude/Wilson loop is divided into several different contributions with very precise physical meaning. It would be very interesting to try to isolate them at an earlier stage (already at the Integrand level if possible) hence simplifying their computation.}\,\footnote{The equivalence between the super Wilson loop and the scattering amplitude was proven at the level of the loop integrand \cite{Skinner,Simon}. These observables however are UV/IR divergent and need regularization. A proper regularization of the super Wilson loop under which the duality with amplitudes stays untouched is not yet known; for example, dimensional regularization does not seem to work \cite{super2}. Here we assume that such a regularized object exists and we read the results from the known scattering amplitudes data. On the other hand, the null correlation function approach of \cite{Alday:2010zy} leads to basically the same super Wilson loop while their regularization is currently under better control. For us, this is not really important; all we care is that a picture with the two null lines sourcing the flux tube exists. Hence, the logic of the paper will not be effected is we replace any reference to the ``Wilson loop" by ``null correlation function". We thank E. Sokatchev for discussions on this point.
} A one loop MHV amplitude is given by a NPWL one loop computation with a gluon stretching from any edge to any other edge of the polygon. Similarly, some tree level NMHV component are given by the propagation of a scalar from one edge to another edge which clearly resembles the one loop computation. A N$^2$MHV amplitude might be described by two such exchanges, resembling therefore a two loop computation and so on, see figure \ref{examples}. Hence a plethora of multi-particle flux tube excitations abounds in the description of N$^k$MHV amplitudes, already at tree level and one loop! 

To be able to analyse this very exciting data through the OPE lens, we need to understand what are the new ingredients in the OPE for supersymmetric NPWL. This is the purpose of the current paper. We will mostly focus on Next-to-Maximally Helicity Violating (NMHV) amplitudes; we will consider N$^k$MHV amplitudes and the multi-particle contributions elsewhere. We will show how to bootstrap the full result for all six particle one loop NMHV amplitudes \cite{addpapers0,addpapers1,freddy} from the tree level data \cite{Drummond:2008cr}. We will also present some new all loop predictions for these amplitudes.

\begin{figure}[t]
\centering
\def\svgwidth{16cm}
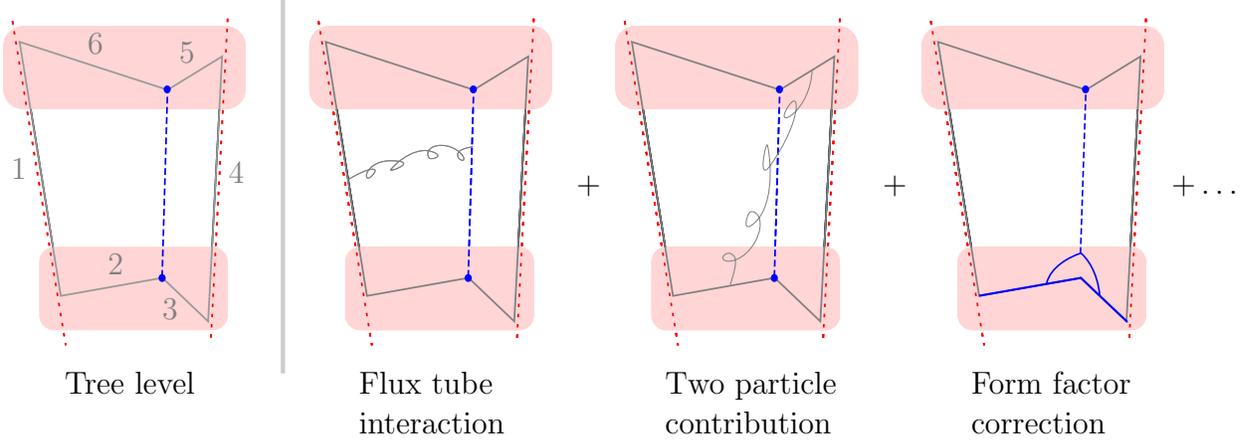
\caption{\small $\eta_2\eta_3\eta_5\eta_6$ component at tree level (left) and at one loop level (right). At one loop there are three contributions that kick in as indicated in the figure. From the OPE point of view the interaction with the flux tube is under control since we can compute the energy of the flux tube excitations exactly using integrability \cite{Benjamin}.}\label{niceex}
\end{figure}

This paper is organized as follows. In section \ref{sec2} we will start by studying the tree level NMHV ratio function. We will explain how we can interpret known results from the OPE point of view. More precisely, we will re-write the results for six gluons as an expansion of particles flowing in the flux tube mentioned above. These expressions are the seeds of the higher loop computations. At one loop there are three types of corrections to the ratio function. Corrections to the energy of the particles, corrections to the form factor and a new two particle contribution, see figure \ref{niceex}. Due to integrability, the correction to the energy of the excitations is under full control. That correction captures what is known as the OPE discontinuity of the amplitude. We will be able to bootstrap the full one loop results from their discontinuities. To compute all discontinuities of the result, we will first compute a restrictive set of discontinuities that have a clear OPE interpretation. All other discontinuities will be obtain from these restrictive set using SUSY Ward identities as explained in section \ref{sec3}. In other words, the role of channel duality is played by the Ward identities.\footnote{One other nice outcome of the Ward identities is that they will allow us to bootstrap all one loop amplitudes. Curiously, this even works for components which vanish at tree level and hence start at one loop only.} In section \ref{sec4} we show how the OPE for NMHV loops can be used to bootstrap the one loop ratio function and in section \ref{sec5} we present some new all loop predictions for the NMHV ratio function. 

\section{Tree Level} \la{sec2}
The observable we will study in this paper is the ratio function for NMHV amplitudes \cite{Drummond:2008vq}. In this section we study this observable at tree level which as we explain below is the seed to the higher loops. It is obtained by extracting components from the NMHV generating functional \cite{Drummond:2008vq,Drummond2,Mason:2010yk}
\beq\la{genNMHV}
\mathcal{R}_n^{\text{NMHV tree}} ={\mathcal{A}_n^{\text{NMHV tree}} \over \mathcal{A}_n^{\text{MHV tree}}  }=\sum_{1<i<j<n-2} [1,i,i+1,j,j+2]
\eeq
where 
the five bracket is the so called R-invariant given by 
\beq
\qquad [a,b,c,d,e]\equiv \frac{\delta^{0|4}\(\eta_{a} \<bcde\>+\eta_{b} \<cdea\>+\eta_{c} \<deab\>+\eta_{d} \<eabc\>+\eta_{e} \<abcd\>\) }{\<abcd\>\<bcde\>\<cdea\>\<deab\>\<eabc\>} \,. \la{Rinvariant}
\eeq
and $\<abcd\>\equiv \epsilon_{ABCD} Z_{a}^AZ_{b}^BZ_{c}^CZ_{d}^D$  is the determinant of four dual momentum twistors. We assume the reader is familiar with these formulae (for nice reviews see \cite{daveR,freddy,lance,school} and references therein). The main purpose of writing them down was to set the notation used throughout the paper. Lorentz, spinor and R-charge indices are always omitted unless their placement is not obvious. As a final disclaimer, we also assume acquaintance with the basic OPE ideas spelled out in \cite{OPEpaper}, see also \cite{bootstraping,Hexagonpaper,heptagon}.\footnote{
In particular, see \cite{OPEpaper} for more details on the concepts of flux tube, flux tube excitations and their anomalous dimensions, reference square, $R_\tau\times R_\sigma\times SO(2)_\phi$ symmetries of the reference square, OPE channel defined by two null lines, $SL(2)_{\tau}$ symmetries of the two null lines and also meaning of the parametrization variables $\tau,\sigma,\phi$ and their conjugate charges $E,p,m$. For the idea of the $SL(2)_{\tau}$ Casimir operator acting on part of a null polygon see sections 3.5 and 5.2 in \cite{Hexagonpaper} and also \cite{heptagon}. For more details on the flux tube anomalous dimensions see \cite{Benjamin,bootstraping,Hexagonpaper}.} In this paper we will focus mostly on the new features of the OPE which arise when considering supersymmetric loops.

For simplicity, in this paper, we will consider  NMHV amplitudes with $n=6$ particles. That is we will consider the Super Null Hexagon Wilson Loop.  The generalization to more particles can be done following \cite{heptagon}.

The OPE description involves thinking of particles being produced in a region of the polygon, propagating in a gluonic flux tube and being absorbed at some other region of the polygon. Hence, most symmetries of the scattering ampltiudes, in particular ciclicity and supersymmetry, are not manifest. In fact, when analyzing amplitudes from the OPE point of view, we found it sensible to \textit{pick components} instead of considering the whole generating function (\ref{genNMHV}) at once. Let us start with a simple example, the $\eta_2 \eta_3 \eta_5 \eta_6$ component 
\beq
\left. \mathcal{R}_6^{\text{NMHV tree}}  \right|_{\eta_2 \eta_3 \eta_5 \eta_6} \equiv  \mathcal{R}^{(2356)}   = \frac{1}{\<2356\>} \la{firstexample} \,,
\eeq
and  analyze it from the OPE channel where the two null lines of the expansion coincide with edge 1 and 4 (see figure \ref{niceex}.a).

An important new conceptual novelty one encounters when dealing with objects like (\ref{firstexample}) is that these carry helicity weights and are therefore not pure functions of conformal cross ratios. That is, the momentum twistors live in projective space and (\ref{firstexample}) is not a c-number. It only becomes a number, measurable in a collider, when contracted with the corresponding polarization vectors.

To apply the OPE procedure to these helicity carrying objects, we will first fix an helicity gauge by picking some momentum twistors $\widetilde Z_a$  for some initial reference hexagon. 
We then generate a three parameter family of momentum twistors $Z(\tau,\sigma,\phi)$ by acting with the three conformal symmetries of the reference square $R_{\tau}\times R_{\sigma}\times SO(2)_{\phi}$ 
on the bottom twistors, that is 
\beq
Z_{a}=\left\{ \begin{array}{ll}
M(\tau,\sigma,\phi)\cdot \widetilde Z_{a} & \text{, for } a=2,3\\
 \widetilde Z_a  &\text{, for }  a\neq 2,3 \end{array}\right. \la{ZZt}
\eeq
where $M(\tau,\sigma,\phi)$ is the $R_{\tau}\times R_\sigma
\times SO(2)_\phi$ transformation mentioned above.\footnote{To be more precise, one acts on the bottom cusps. These are located at $\widetilde Z_1\wedge\widetilde Z_2,\ \widetilde Z_2\wedge\widetilde Z_3$ and $\widetilde Z_3\wedge\widetilde Z_4$. That is, they also depends on $\widetilde Z_1$ and $\widetilde Z_4$. However, $\widetilde Z_1$ and $\widetilde Z_4$ only transform under $R_\tau\times R_\sigma\times SO(2)_\phi$ by an overall rescaling (see (\ref{twistors})). For the channel defined by edges 1 and 4, we will only consider those NMHV amplitudes that do not carry $\eta_1$ and $\eta_4$ helicity weights. These are invariant under overall rescalings of $\widetilde Z_1$ and $\widetilde Z_4$.} 
For the  choice of reference square used in \cite{Hexagonpaper}, see also appendix \ref{details}, we find  
\beq\la{ffunc}
\cR^{(2356)}[Z_i]=\frac{1}{4[\cosh(\sigma)\cosh(\tau)+\cos(\phi)]}
\eeq
The overall normalization of the right hand side of (\ref{ffunc}) depends on our choice of helicity gauge $\widetilde Z_{a}$, see (\ref{twistors}). We would like to express (\ref{ffunc}) as a sum of flux tube excitations propagating from the bottom to the top. These can be classified into $SL(2)$ primaries and descendants. The NMHV amplitude (\ref{firstexample}) is obtained from a scalar propagator between cusp $Z_2\wedge Z_3$ and $Z_5\wedge Z_6$. Therefore, all primaries are expected to be scalars.  All such primaries are 
\beq\la{scalarexcitations}
\partial_z^m \phi_{AB} \,, \qquad \partial_{\bar z}^m \phi_{AB} \,,\quad \text{where  } \,m=0,1,2,\dots
\eeq  
Here, $\d_z,\d_{\bar z}$ are spacetime derivatives in the transverse directions to the reference square, carrying charge $\pm1$ under $SO(2)_\phi$. 
There is a very simple check one can do to confirm that indeed, (\ref{ffunc}) has a  decomposition in these excitations and their conformal descendents. The twist of these excitations is $E=\Delta-S=1+|m|$. The corresponding $SL(2)$ Casimir is $C=E(E-2)=(m^2-1)$.\footnote{Our convention for the SL(2) Casimir here differ by a factor of 4 with respect to \cite{bootstraping,Hexagonpaper}.} In particular, it is independent of the sign of $m$ and therefore its action on  excitations (\ref{scalarexcitations}) is given by the differential operator $-(\d_\phi^2+1)$. On the other hand, the $SL(2)$ Casimir can be represented as a second order differential operator 
\beq
{\mathbb C}=-J_1^2+J_2^2-J_3^2 \nn
\eeq
acting on the bottom of the polygon. Its action on the ratio function can be computed as\footnote{In the paremetrization presented in the appendix, $ J_1=\mathbb{I}_{2\times 2} \otimes \sigma_1$, $ J_2=i\mathbb{I}_{2\times 2} \otimes \sigma_2$, $ J_3=\mathbb{I}_{2\times 2} \otimes \sigma_3$  where the $\sigma_a$'s are Pauli matrices. \la{footnote} Note that $Z_\text{right}$ and $Z_\text{left}$ are invariant under the $SL(2)_\tau$ action and therefore the result is the same whether we act on them or not.}
\beq
\mathbb{C}\circ \cR [Z_i] = \sum_{a=1}^3(-1)^{a+1} \left.\partial_{\theta}^2\, \cR \[Z_\text{right},\,\{Z_{top}\},\,Z_\text{left},\, \{e^{i\theta J_a} \circ Z_{bot}\}\]\right|_{\theta=0} \nn 
\eeq
We conclude that the ratio function has the expected decomposition if and only if it is annihilated by the second order differential operator
\beq\la{projector}
{\mathbb{ P}}={\mathbb C}+(\d_\phi^2+1)\ .\nn
\eeq 
Indeed, we find\footnote{See also \cite{school}} $${\mathbb{ P}}\circ \cR^{(2 3 5 6)}=0\,.$$ This is first non-trivial evidence for the OPE expansion of super loop! Let us check another component where we also expect scalar excitations to be flowing, 
\beq
\cR^{(2256)}[Z_i]   =\frac{\<3561\>}{\<2356\>\<1256\>}
=    -\frac{\text{csch}(\tau )e^{\sigma +\phi }+\coth (\tau )}{
   4\cosh (\sigma ) \cosh (\tau )+4\cosh (\phi )} \nn
\eeq
Once again we find 
$${\mathbb{P}}\,\circ \,\cR^{(2 2 5 6)}[Z_i]=0\,.$$ 
It is useful to convert the action of $\mathbb C$ on $\cR^{(ijkl)}[Z_i(\tau,\sigma,\phi)]$ into a differential operator $\mathbb C^{(ijkl)}$ acting on $\sigma, \tau, \phi$. We find 
\beqa\la{diffoperator}
{\mathbb C}^{(2356)}&=&\[\partial_{\tau}^2+2\coth(2\tau)\partial_{\tau}+{\rm sech}^2(\tau) \partial_{\sigma}^2\] \\
\la{diffoperator2}
{\mathbb C}^{(2256)}&=&\[\partial_{\tau}^2+2\coth(2\tau)\partial_{\tau}+{\rm sech}^2(\tau) \partial_{\sigma}^2\]+\[{\rm sech}^2(\tau)(1-2\d_\sigma)-\text{csch}^2(\tau)\] 
\eeqa
Note that the operator $\mathbb C$, when converted into a differential operator acting on $\sigma, \tau, \phi$, takes different forms when acting on the $\eta_2 \eta_3 \eta_5 \eta_6$ component (\ref{diffoperator}) and on the $\eta_2 \eta_2 \eta_5 \eta_6$ component (\ref{diffoperator2}). This arises because we are dealing with functions carrying helicity weight. The reader who is not puzzled with this statement should probably skip the next paragraph (in smaller font) in a first reading.

\begin{itemize}
\item[]
\footnotesize
To understand why the differential operator $C^{(ijkl)}$ takes different forms acting on different ratio function components lets us be more general and consider an arbitrary conformal invariant function ${\cal H}^{(ijkl)}[Z_a]$ carrying helicity wight $\eta_i\eta_j\eta_k\eta_l$.  Ratio function components are examples of such functions. Even though ${\cal H}^{(ijkl)}[Z_a]$ is conformal invariant, it is \textit{not} a function of the conformal cross ratios only. Instead, it transforms under overall resealing of the twistors $\{Z_i,Z_j,Z_k,Z_l\}$. The ratio $h^{(ijkl)}= {\cal H}^{(ijkl)}[Z]/\cR^{(ijkl)}[Z]$ however, is independent of the helicity gauge chosen and is therefore a function of the conformal cross ratios $u_i$ only. 
In other words,
\beq
\mathcal{H}^{(ijkl)}[Z_a]=h^{(ijkl)}(u_1,u_2,u_3) \mathcal{R}^{(ijkl)}[Z_a] \la{ratio}
\eeq
where (\ref{ZZt}) and $u_i$'s are the cross ratios computed from the transformed $Z$'s (\ref{theus}). The parametrization of $u_1,u_2$ and $u_3$ in terms of $\tau,\sigma$ and $\phi$ is given in the Appendix, see (\ref{ccr}).  (Implicitly, we assumed that the tree level ratio function component $\cR^{(ijkl)}\ne0$; if $\cR^{(ijkl)}=0$ one can use an arbitrary reference conformal invariant funcrion carrying $(ijkl)$ weight.)
Under an $SL(2)$ transformation $g$ acting on the bottom twistors,
\beqa
\mathcal{H}^{(ijkl)}[Z] \to h^{(ijkl)}(g\circ \tau,g\circ \sigma,g\circ \phi) \mathcal{R}^{(ijkl)}[ Z_1,g\cdot Z_2,g\cdot Z_3,Z_4, Z_5, Z_6] \la{transf}
\eeqa
where $g\circ\tau$ and $g\circ\sigma$ stands for the action on $\tau$ and $\sigma$ dictated from the transformation of  the cross-ratios (\ref{inversemap}), see appendix. 
This action is of course universal. However, the last term in (\ref{transf}) has helicity weight and transforms differently for different components. That transformation gives a non trivial contribution to the differential operator $\mathbb C^{(ijkl)}$ and is the source for the difference between ${\mathbb C}^{(2356)}$ and ${\mathbb C}^{(2256)}$ for example. A more thorough discussion of this point is given in appendix \ref{A2}. 
\end{itemize}
\normalsize

We will now decompose (\ref{ffunc}) in the scalar excitations propagating from the bottom to the top. The propagation of a conformal primary and its conformal descendants are conveniently packed together in a conformal block. These are functions of the form (\ref{ratio}) ${\cal H}^{(ijkl)}=e^{ip\sigma}\cF^{(ijkl)}_{{E},p}(\tau)$ that solve the differential equation\footnote{We use a slightly different notation here compared to \cite{Hexagonpaper,heptagon}. Namely, $\left.\cF^{(ijkl)}_{{E},p}(\tau)\right|_{here}=\left.\cF^{(ijkl)}_{{E\over2},p}(\tau)\right|_{there}$. The conformal cross ratio $\phi$ do not enter the equation because it does not transform under the $SL(2)$ transformation $g\circ\phi=\phi$.}
\beq\la{cblockeq}
{\mathbb C}^{(ijkl)}\circ \[e^{ip\sigma}\cF^{(ijkl)}_{{E},p}(\tau)\]=E(E-2)\[e^{ip\sigma}\cF^{(ijkl)}_{{E},p}(\tau)\]
\eeq
The relevant solution of that equation with $(ijkl)=(2356)$ is given in \cite{Hexagonpaper} and reads
\beq\la{confF}
\mathcal{F}_{E,p}^{(2356)}(\tau)={\rm sech}^{E} (\tau)\ _2F_1\[\frac{E-ip}{2}, \frac{E+ip}{2} ,E,{\rm sech}^{2}(\tau)\]\,.
\eeq
We find that
\beqa
\cR^{(2 3 5 6)}[Z]
=\sum_{m=-\infty}^{\infty} \int \frac{dp}{2\pi} e^{i m \phi-i p \sigma} \mathcal{C}_{m}^{(2356)}(p) \mathcal{F}_{{|m|+1},p}^{(2356)}(\tau) \la{2356exp}\,,
\eeqa
where the form factors are given by
\beq
\mathcal{C}^{(2356)}_m(p)= {1\over4}(-1)^m B\(\frac{|m|+1+ip}{2},\frac{|m|+1-ip}{2}\) \la{C2356}\,.
\eeq
For our other example we find exactly the same expansion with extremely similar blocks and form factors, (the two minor differences are colored) 
\beqa
\la{2256exp} \cR^{(2 2 5 6)}[Z]
&=&\sum_{m=-\infty}^{\infty} \int \frac{dp}{2\pi} e^{i m \phi-i p \sigma} \mathcal{C}_{m}^{(2256)}(p) \mathcal{F}_{{|m|+1},p}^{(2256)}(\tau) \nn\\
 \mathcal{F}_{{E},p}^{(2256)}(\tau)& =&\tanh(\tau) \,{\rm sech}^{E} (\tau)\ _2F_1\[\frac{E-ip}{2}, \frac{E+ip\color{magenta}+2\color{black}}{2} ,E,{\rm sech}^{2}(\tau)\]  \nn\\
\mathcal{C}^{(2256)}_m(p)&=&\mathcal{C}^{(2356)}_m(p\color{magenta}-2i\, \delta_{m\ge 0}\color{black}) \,. \nn   
\eeqa
It is very curious to notice that these expansions are shockingly similar to each other and also to the expansions arising for the MHV hexagon; compare for example (\ref{ffunc}), (\ref{2356exp}) with equations (75),(76) from \cite{Hexagonpaper}. The form factors (\ref{C2356}) are exactly the same as those arising in the expansion of $\Box \,r_{U(1)}$. This seems to indicate that there is some sort of hidden symmetry relating them. In fact, this statement is even more reinforced when we consider other tree level components with a good OPE expansion in the channel defined by the edges $1$ and $4$, see  figure \ref{examples} for examples.
\begin{figure}[t]
\centering
\def\svgwidth{15cm}
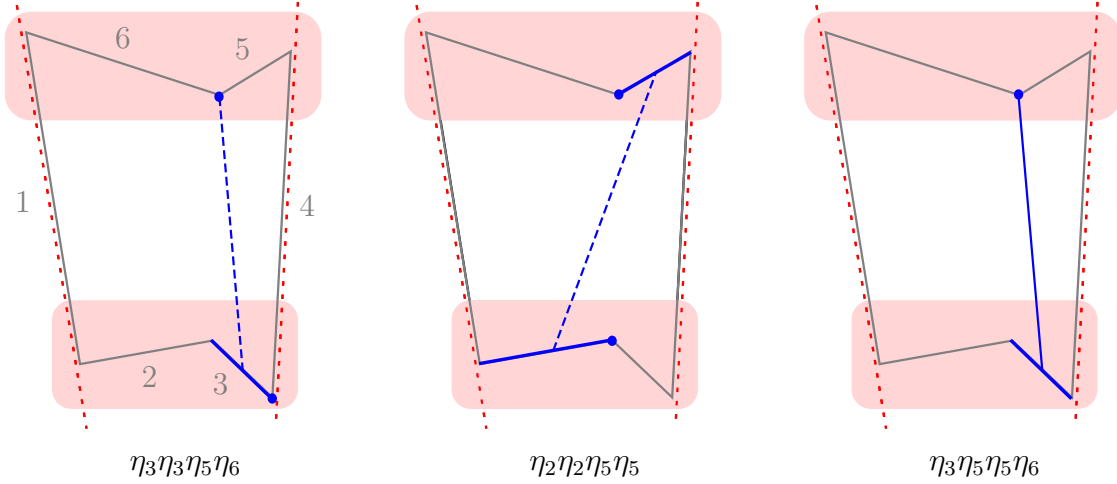
\caption{\small Some components with a good OPE expansion in the channel defined by the null edges 1 and 4. The first two are given by scalars propagating from bottom to top and the third is described by propagation of fermions. }\label{examples2}
\end{figure}
In terms of momentum twistors these component look very different from each other. However, when re-written as a decomposition over flux tube excitations, we found that they all look roughly the same up to very minor redefinitions of the conformal blocks and form factors which typically just involve some simple shifts in $p$ and $m$! These similarities are partially a consequence of supersymmetry which tells us that only five NMHV components are actually independent. All other components can be obtained from these ones through SUSY Ward identities \cite{Elvang}. There ought to be a more supersymmetric way of packaging all these expansions together into some sort of SUSY blocks and/or SUSY form factors.

\begin{figure}[t]
\centering
\def\svgwidth{16cm}
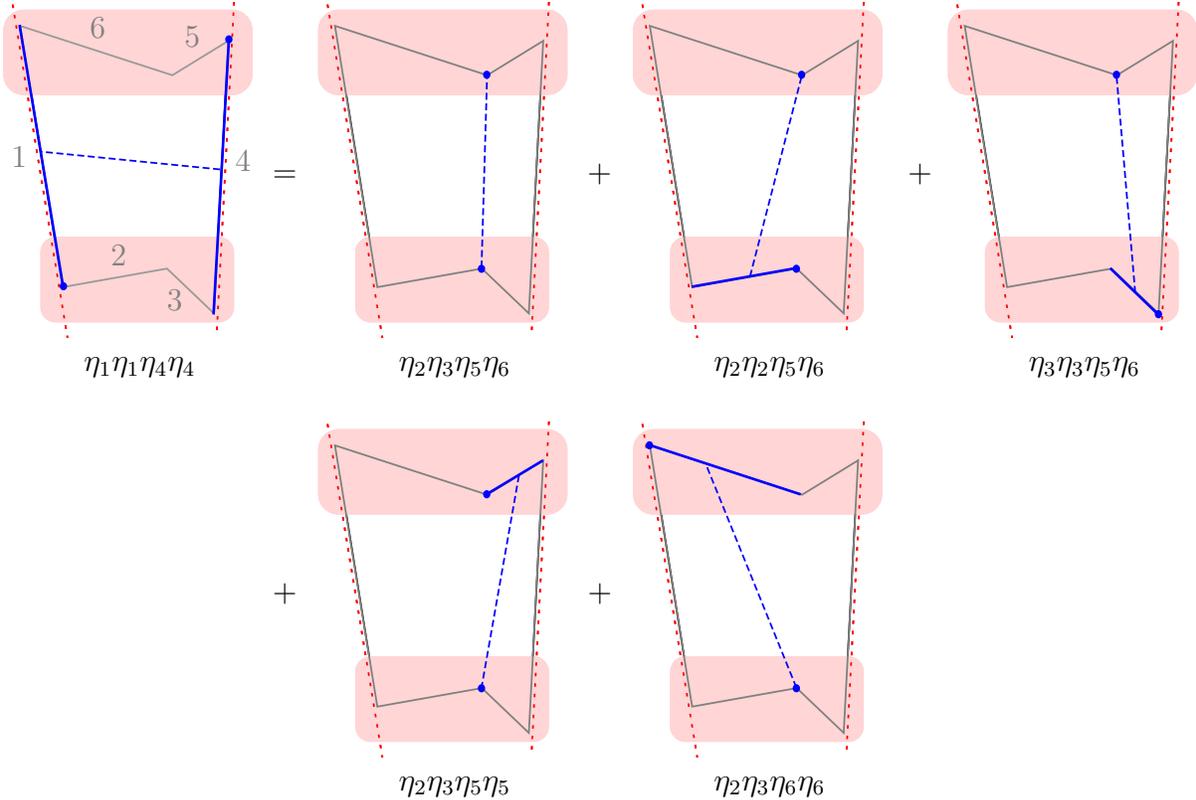
\caption{\small Ward Identities allow us to relate any six point scattering amplitude to a base of five components. In particular this allows us to access the behavior of any amplitude in any OPE channel. In the figure the component $\eta_1\eta_1\eta_4\eta_4$, which clearly does not have a natural expansion in the channel defined by the null edges 1 and 4 is expanded in terms of five components that do admit a neat expansion in this channel. This can be done at any loop order. Ward identities replace channel duality in this problem.
}\label{wardid}
\end{figure}

\section{SUSY} \la{sec3}
In this section we will review how different components of the NMHV ratio function are related to each other through supersymmetry \cite{Elvang}. We will also study some implications of cyclicity and parity. 

As explained in the previous section, when considering the OPE expansion in some channel, there are particularly nice components which admit a simple expansion in that channel. For example, the amplitudes in figure \ref{examples2} have a nice expansion in the OPE channel defined by the null edges 1 and 4. Naively, these components are tricky to expand in other channels. On the other hand, to bootstrap the amplitudes at higher loops we need to be able to expand any component in any possible channel! SUSY Ward identities provide a neat way out. Using them we can rotate between components are infer the OPE expansion of some components in channels where naively that OPE expansion is hard to predict directly, see figure \ref{wardid} for one such example. In other words, they replace/implement channel duality. This will be crucial when bootstraping one loop amplitudes from  their OPE discontinuities in all possible channels. 

The generating functional of NMHV ratio functions  is superconformal and  dual superconformal. Hence it ought to be given by a linear combination of R-invariants (\ref{Rinvariant}). For the hexagon only five are independent \cite{Elvang}. Without loss of generality we can write
\beqa
 \mathcal{R}^{\text{NMHV}}\!= [6\, 2\, 3\, 4\, 5]\, \mathfrak{A} + [6\,1\,2\,3\,5] \, \mathfrak{B}    +[6\,1\,2\,4\,5]  \, \mathfrak{C} +[6\,1\,2\,3\,4]   \, \mathfrak{D} +[6\,1\,3\,4\,5] \, \mathfrak{E} \la{wardeq} \,.
   \eeqa
%
%
Next, we express $\mathfrak{A},\dots,\mathfrak{E}$ in terms of five components that we choose. For example, a particularly nice choice of components is the one in figure \ref{wardid},
\beq
\cR^{(2356)} \,,  \cR^{(2256)} \,, \, \cR^{(3356)}\,,\,\cR^{(2355)}\,, \, \cR^{(2366)}  \,. \la{base}
\eeq
For example
\beqa
\mathfrak{A}=-\frac{ \langle 1235\rangle  \langle 2345\rangle }{\langle 1234\rangle
   }\cR^{(2355)}-\frac{ \langle 1236\rangle  \langle 2345\rangle }{\langle 1234\rangle
   }\cR^{(2356)} \nn
   \eeqa
and so on (see appendix for explicit expressions). Note that these expressions are valid at any loop order, they are simply encoding SUSY Ward identities \cite{Elvang}. 
Actually, all we need to compute is two components!,  for example the two components $\cR^{(2356)} \,,  \cR^{(2256)}$ considered in the previous section are enough. The other three of our base can then be obtained by simple symmetries of the hexagon, 
\beq
\cR^{(2256)}=\left.\cR^{(3356)}\right|_{\sigma\to -\sigma}=-\left.\cR^{(2355)}\right|_{\phi\to -\phi}=-\left.\cR^{(2366)}\right|_{\phi\to -\phi\,,\,\sigma\to-\sigma}\,\,. \la{ciclicity}
\eeq
We know how to promote all five amplitudes (\ref{base}) in the OPE channel defined by the null edges $1$ and $4$. This was the reason why we chose this particular base.
On the other hand, some other components such as $\cR^{(1144)}$ do not have an obvious expansion in this channel, see figure \ref{wardid}. 
However, given (\ref{wardeq}), we can now write any component as a linear combination of the base of amplitudes (\ref{base}). 
In this way we can study the behavior of $\cR^{(1144)}$ in a channel that was naively inaccessible, see figure \ref{wardid}. 

To summarize: we can study any channel we want, for any component $\cR^{(ijkl)}$ we want, by choosing a base of five components with a good OPE expansion in that channel and by writing $\cR^{(ijkl)}$ in terms of that base. 

Using these tricks we can easily compute the one loop discontinuities of all NMHV components in all possible channels. With this information we can bootstrap the full one loop ratio function. This is the subject of the next section.

\section{One Loop} \la{sec4}
One loop amplitudes will have discontinuities at $u_a=0$. In the Euclidean sheet we do not expect any other discontinuities \cite{Hexagonpaper}. The three discontinuities $D_a$ can be computed by the OPE in the three possible channels \cite{Hexagonpaper}. For MHV this suffices to determine the full two loop remainder function. In this section we show that the same is true for the NMHV ratio function at one loop. 

\subsection*{From Tree Level to One Loop OPE Discontinuities}
Consider for example the two components studied in section \ref{sec2}. Their OPE tree level decomposition is given in formulae (\ref{2356exp}) and (\ref{2256exp}). We obtain the discontinuity in the  $u_2=\cosh^{-2}\tau$  channel by simply dressing these expansions by the anomalous dimensions of the scalar primaries flowing in the tube \cite{Hexagonpaper}. That is 
\beq
D_2^{(ijkl)} =\sum_{m=-\infty}^{\infty} \int \frac{dp}{2\pi} \,e^{i m \phi-i p \sigma} \,\mathcal{C}_{m}^{(ijkl)}(p)\, \mathcal{F}_{{|m|+1},p}^{(ijkl)}(\tau) \color{blue}\,\gamma_{1+|m|}(p) \color{black}\,, \nn 
\eeq
where $(ijkl)\in\{(2356),(2256)\}$ and \cite{Benjamin,Hexagonpaper,bootstraping,otherpapers}
\beq\la{gammam}
\gamma_{1+|m|}(p)=2g^2 \[\psi\(\frac{1+|m|+ip}{2}\)+\psi\(\frac{1+|m|-ip}{2}\)-2\psi(1)\] \,.
\eeq
Preforming these sums we get the corresponding OPE discontinuities of the one loop results
\beqa
D_2^{(2356)} &=& \mathcal{R}_{\text{tree}}^{(2356)} \,2 \log\( \frac{u_1 u_3}{1-u_2}\) \nn\\
D_2^{(2256)} &=& \mathcal{R}_{\text{tree}}^{(2256)} \,2\[\log\( \frac{u_1}{1-u_2}\)-\frac{u_3\log(u_3)}{1-u_3} \frac{-1-u_1+u_2+u_3+\sqrt{u_1u_2u_3}\(\mu+\frac{1}{\mu}\)}{+1-u_1+u_2+u_3+\sqrt{u_1u_2u_3}\(\mu-\frac{1}{\mu}\)}\] \nn
\eeqa
Here,  $(\mu=e^{i \phi})$ and we have factored out the tree level results in order to express the result as a function of conformal cross ratios only, without the helicity weights ambiguities discussed in section \ref{sec2}. 
We can now get the other three components in (\ref{base})  from simple parity and ciclicity symmetries (\ref{ciclicity})
\beqa
D_2^{(3356)} &=&\mathcal{R}_{\text{tree}}^{(2256)} \,2\[\log\( \frac{u_3}{1-u_2}\)-\frac{u_1\log(u_1)}{1-u_1} \frac{-1-u_3+u_2+u_1+\sqrt{u_1u_2u_3}\(\mu+\frac{1}{\mu}\)}{+1-u_3+u_2+u_1+\sqrt{u_1u_2u_3}\(\mu-\frac{1}{\mu}\)}\] \nn \\
D_2^{(2355)} &=& \mathcal{R}_{\text{tree}}^{(2256)} \,2\[\log\( \frac{u_1}{1-u_2}\)-\frac{u_3\log(u_3)}{1-u_3} \frac{-1-u_1+u_2+u_3-\sqrt{u_1u_2u_3}\(\mu+\frac{1}{\mu}\)}{+1-u_1+u_2+u_3+\sqrt{u_1u_2u_3}\(\mu-\frac{1}{\mu}\)}\] \nn\\
D_2^{(2366)} &=& \mathcal{R}_{\text{tree}}^{(2256)} \,2\[\log\( \frac{u_3}{1-u_2}\)-\frac{u_1\log(u_1)}{1-u_1} \frac{-1-u_3+u_2+u_1-\sqrt{u_1u_2u_3}\(\mu+\frac{1}{\mu}\)}{+1-u_3+u_2+u_1+\sqrt{u_1u_2u_3}\(\mu-\frac{1}{\mu}\)}\] \nn
\eeqa

\subsection*{Discontinuities from Ward identities, Ciclicity and Parity}
We have just computed discontinuities of the chosen base of five amplitudes (\ref{base}) around $u_2=0$. Using the decomposition of the ratio function in terms of these five amplitudes (\ref{ward}) we can now read the discontinuity of \textit{any} component in this channel. We simply need to put together the five discontinuities computed above with the correct prefactors as predicted by (\ref{wardeq}).
 For example, for the example in figure \ref{wardid} we find
\beqa
D_2^{(1144)} &=&\mathcal{R}_{\text{tree}}^{(1144)} \[ - 2 \log(1-u_2) + \frac{2u_2(u_1-u_3)\log(u_3/u_1)}{(1-u_2)(1-u_1-u_3)} \] \nn \,.
\eeqa
This was one of the \textit{tricky} components from the $u_2=0$ OPE channel point of view, see figure \ref{wardid}. 
Using cyclicity of the generating function, we can now read the discontinuities of any amplitude in the other two channels as well. These are the discontinuities around $u_1=0$ and $u_3=0$. For example, 
\beq
\frac{D_1^{(2356)}}{\cR^{(2356)}}=\(\frac{D_2^{(1245)}}{\cR^{(1245)}}\)_{u_i \to u_{i-1}} \,,  \qquad \frac{D_3^{(2356)}}{\cR^{(2356)}}=\(\frac{D_2^{(1245)}}{\cR^{(1245)}}\)_{u_i \to u_{i+1}} \nn
\eeq
etcetera.\footnote{Note that in some cases like the component (3556), the OPE discontinuity is associated with the corrected propagation of a fermion instead of a scalar, see figure \ref{examples2}. The dependence of the fermion anomalous dimension on the $SO(2)_\phi$ charge $m$ is different from the scalar one by a 1/2 shift: $\gamma_{1+|m+1/2|}$ instead of $\gamma_{1+|m|}$. Still, the discontinuities for the two cases are simply related to each other. The way it comes about is that in the fermionic case, the charge $m$ runs over half integer values so after factoring out an $e^{-\phi/2}$, the dependence on the $SO(2)_\phi$ charge is the same.}
\subsection*{Bootstrapping the Full Amplitude From its Discontinuities}

It is now a very simple task to reconstruct the full function for our base from those discontinuities. One scientific way to proceed would be to use symbols \cite{Gon,Hexagonpaper}. 
This works perfectly. On the other hand, we are dealing with a very simple expressions of transcendentally degree two. Hence, 
will proceed in a more pedestrian and less fancy way and illustrate, on a simple example, how to obtain the form of the full result from the knowledge of its three OPE discontinuities.  

Let us consider the particularly simple component $\cR^{(2356)}$; the other cases are roughly of the same complexity. Its three discontinuities can be derived from the OPE as explained above. One obtains
\beqa
D_1^{(2356)} &=&2 \,\mathcal{R}_{\text{tree}}^{(2356)} \,\Big[ \log u_2- \log u_3- \log(1-u_1) \Big]\la{d1}\\
D_2^{(2356)} &=&2 \,\mathcal{R}_{\text{tree}}^{(2356)} \,\Big[ \log u_1 + \log u_3- \log(1-u_2) \Big] \la{d2} \\
D_3^{(2356)} &=&2 \,\mathcal{R}_{\text{tree}}^{(2356)} \,\Big[ \log u_2- \log u_1- \log(1-u_3) \Big]\la{d3}
\eeqa
Consider first the discontinuity (\ref{d1}) in the channel $u_1=0$. What function has $ \log u_2- \log u_3- \log(1-u_1)$ as a discontinuity? There are two obvious options:
\beq
\log u_1 \Big(\log u_2- \log u_3-  \log(1-u_1) \Big)+\dots \qquad \text{or} \qquad \log u_1 \Big( \log u_2-  \log u_3 \Big)-{\rm Li}_2(1-u_1) +\dots \nn
\eeq
where $\dots$ stands for extra terms without any discontinuity around $u_1=0$. It is clear that we should pick the second option because the first option has a discontinuity at $u_1=1$ and  the only allowed discontinuities in the Euclidean sheet are at $u_a=0$ and $u_a=\infty$, when two cusps become null separated. Hence
\beqa
\cR_{\text{one loop}}^{(2356)} &=&2 \,\mathcal{R}_{\text{tree}}^{(2356)} \,\Big[ \log u_1\log u_2- \log u_1\log u_3-{\rm Li}_2(1-u_1)  +\dots \Big]\la{sofar}
\eeqa
Next we look at the second discontinuity, in the channel $u_2=0$. The first term in (\ref{d2}) is already accounted by the first term in (\ref{sofar}); the second term leads us an extra $\log u_2 \log u_3$; the last term gives rise to another ${\rm Li}_2$ by the same reason as argued above. Hence 
\beqa
\cR_{\text{one loop}}^{(2356)} &=&2 \,\mathcal{R}_{\text{tree}}^{(2356)} \,\Big[ \log u_2\log (u_1 u_3)- \log u_1\log u_3-{\rm Li}_2(1-u_1)  -{\rm Li}_2(1-u_2)  +\dots \Big] \nn
\eeqa
We only need to consider the last discontinuity (\ref{d3}). We easily see that to reproduce it we only need to add another polylogarithm to the result. That is,  
\beqa
\cR_{\text{one loop}}^{(2356)} &=&2 \,\mathcal{R}_{\text{tree}}^{(2356)} \,\Big[ \log u_2\log (u_1 u_3)- \log u_1\log u_3- \sum_{a=1}^3 {\rm Li}_2(1-u_a)   + c \Big] \nn
\eeqa
where $c$ is a constant of transcendentally degree two which we will soon fix using collinear limits. This is the prediction of the OPE approach.\footnote{Implicitly we also assume that at a given loop order the transcendentality degree of the result is uniform. Empirically, this seems to be a true property of $\mathcal{N}=4$ SYM.} We can now compare it with the known results computed in the literature by more direct methods \cite{addpapers0,addpapers1,freddy}. We find a perfect match with the OPE prediction.

All other components can be derived in the same way. Due to the Ward identities reviewed in the previous section, all we need is four more components: with a base of five amplitudes we can extract any component through (\ref{wardeq}). Furthermore, as also explained above, only one out of these four is really new, the other ones are trivially obtained from this one (\ref{ciclicity}).  Proceeding as above we derive that missing component,  
\beqa
\cR^{(2256)}_{\text{one loop}}=2\,\cR^{(2256)}_{\text{tree}} \!\!\!\!\!\!\!\!\!\!&&\Big[ \log(u_1)\log(u_2)+c' +\sum_{a=1}^3 {\rm Li}_2(1-u_a) \nn \\
&&+\,\,\frac{\(u_1+u_2+u_3-2u_1 u_3-1+\(\displaystyle \mu+\frac{1}{\mu}\)\sqrt{\,u_1u_2u_3}\)\log(u_3)\log\(\displaystyle\frac{u_2}{u_1}\)}{2(1-u_1)(1-u_3)}\Big] \nn
\eeqa
We are done concerning one-loop except for the two constants $c$ and $c'$ which we did not fix yet. These can be easily fixed from collinear limits to be $c=c'=-\pi^2/3$. 

\begin{itemize}
\item[]
\small The idea is that many components like $\cR_{\text{one loop}}^{(1144)}/\cR_{\text{tree}}^{(1144)} $ or  $\cR_{\text{one loop}}^{(1134)}/\cR_{\text{tree}}^{(1134)}$ ought to vanish in the $\tau\to \infty$ limit. That is, any component that do not involve the edges $\{2,3\}$ or the edges $\{5,6\}$ will approach the NMHV pentagon in the $\tau\to\infty$ collinear limit. The NMHV pentagon is an $\overline{\text{MVH}}$ amplitude so that the ratio function does not receive loop corrections. These components can be written in terms of the amplitudes derived above. In this way one concludes that
\beq
c=c'=-\frac{\pi^2}{3} \,. \nn
\eeq
\normalsize
\end{itemize}
This concludes the derivation of all one loop NMHV six point amplitudes from the OPE point of view. Note in particular that using the Ward identities we also get those one loop NMHV amplitudes that are zero at three level. Next we move to higher loops.

\section{All Loop Predictions} \la{sec5}
Using the OPE expansion we can easily predict an infinite amount of data for higher loop NMHV amplitudes. For example, at $n$ loops we expect \cite{OPEpaper}
\beq
\cR^{(2356)}_{\text{n loops}}= D_i^{(n,0)}(u_1,u_2,u_3)+ D_i^{(n,1)}(u_1,u_2,u_3) \log(u_i)+\dots + D_i^{(n,n)}(u_1,u_2,u_3) \log(u_i)^n\nn
\eeq 
where $D_i^{(n,k)}(u_a)$ have regular expansions in powers of $u_i$ around $u_i=0$. Normally each $D_i^{(n,k)}$ should have an extra index $(2356)$ which we omitted to render the expression lighter. 
In the previous section we showed how to compute the discontinuities $D_i^{(1,1)} \equiv D_i$ in all possible channels and we saw that they are sufficient to reconstruct the full NMHV result at one loop. The \textit{highest discontinuity} $D_i^{(n,n)}$ at $n$ loops is equally simple to get. 
For example \cite{OPEpaper}
\beq
D_2^{\color{purple}(n,n)\color{black}}=\color{purple}\frac{1}{n!} \color{black}\sum_{m=-\infty}^{\infty} \int \frac{dp}{2\pi} \,e^{i m \phi-i p \sigma} \,\mathcal{C}_{m}^{(ijkl)}(p)\, \mathcal{F}_{{|m|+1},p}^{(ijkl)}(\tau) \color{blue}\,\gamma_{|m|+1}(p)^{\color{purple} n }  \color{black} \nn
\eeq
where $\cF,\cC$ and $\gamma$ are given in (\ref{confF}), (\ref{C2356}) and (\ref{gammam}). 
To compute the maximal discontinuity in other channels we use the Ward identities exactly as above. The maximal discontinuities of other components are computed in the same way. 

This contribution is the easiest to extract from the OPE point of view. To bootstrap the full two loop NMHV ratio function one also needs to compute $D_i^{(2,1)}$. That requires a better handling of multi-particle flux tube excitations and is therefore beyond the scope of the present paper.

\section*{Acknowledgments}

We thank N.\ Arkani-Hamed, B.\ Basso, Z.\ Bern, F.\ Cachazo, J.\ Maldacena, J.\ Penedones, D.\ Skinner, E.\ Sokatchev, E.\ Witten,  the participants of the \textit{The Harmony of Scattering Amplitudes} program at KITP for  very useful discussions. We thank J. Bourjaily for a maximally useful \verb"Mathematica" notebook which we used to generate tree level and one loop Next-to-Maximally-Helicity-Violating ratio functions.
The research of AS and PV has been supported in part by the Province of Ontario through ERA grant ER 06-02-293. Research at the Perimeter Institute is supported in part by the Government of Canada through NSERC and by the Province of Ontario through MRI. A.S. and
P.V. would like to thank KITP for warm hospitality. 
This work was partially funded by the research grants PTDC/FIS/099293/2008 and CERN/FP/116358/2010. This research was supported in part by DARPA under Grant No.
HR0011-09-1-0015 and by the National Science Foundation under Grant
No. PHY05-51164.

\appendix


\section{Details on the Hexagon Computation} \la{details}
\subsection{Parametrization and cross-ratios}
As mentioned in the text a family of hexagons is generated by starting with some random twistors $\widetilde Z_i$ and acting with the symmetries of  a reference square on a subset of those twistors, those constituting the \textit{bottom} part of the polygon. A nice parametrization which we use throughout the paper is
\beq\la{twistors}
\(\!\! \begin{array}{c}
Z_1 \\
Z_2 \\
Z_3 \\
Z_4 \\
Z_5 \\
Z_6 
\end{array}\!\!\) =\(\!\! \begin{array}{r}
\widetilde Z_1 \\
M \cdot \widetilde Z_2 \\
M \cdot \widetilde Z_3 \\
\widetilde Z_4 \\
\widetilde Z_5 \\
\widetilde Z_6 
\end{array}\!\!\)=
\left(\!\!
\begin{array}{cccc}
 0 & 1 & 0 & 0 \\
 0 & e^{\frac{\phi }{2}-\sigma } & -e^{\tau -\frac{\phi }{2}} & e^{-\tau
   -\frac{\phi }{2}} \\
 e^{\sigma +\frac{\phi }{2}} & 0 & -e^{\tau -\frac{\phi }{2}} & -e^{-\tau
   -\frac{\phi }{2}} \\
 1 & 0 & 0 & 0 \\
 1 & 0 & 1 & 1 \\
 0 & 1 & 1 & -1
\end{array}
\!\!\right)
\eeq
The left and right null lines correspond to  $\{Z_{\text{left}}=Z_1,\,\hat Z_{\text{left}}=Z_6\wedge Z_1\wedge Z_2\}$ and $\{Z_{\text{right}}=Z_4,\,\hat Z_{\text{right}}=Z_3\wedge Z_4\wedge Z_5\}$. The top and bottom twistors of the reference square are $Z_{\text{top}}=(0,0,1,0)$ and $Z_{\text{bottom}}=(0,0,0,1)$.
The $SL(2)$ group which preserves the two null lines  $\{Z_{\text{left}},\hat Z_{\text{left}}\}$ and $\{Z_{\text{right}},\hat Z_{\text{right}}\}$  acts on the two last slots of the twistors as $\mathbb{I}_{2\times 2}\otimes g_{SL(2)}$, see footnote \ref{footnote}.
A base of cross ratios
\beq\la{theus}
u_1=\frac{\langle 3456\rangle  \langle 6123\rangle }{\langle 3461\rangle 
   \langle 5623\rangle }\,,\qquad u_2=\frac{\langle 1234\rangle  \langle 4561\rangle
   }{\langle 1245\rangle  \langle 3461\rangle }\,,\qquad u_3=\frac{\langle 2345\rangle 
   \langle 5612\rangle }{\langle 2356\rangle  \langle 4512\rangle } 
\eeq
reads  \cite{Hexagonpaper}
\beq
u_1=\frac{e^{\sigma } \sinh (\tau ) \tanh (\tau )}{2 \cosh (\sigma ) \cosh (\tau
   )+2\cosh (\phi )} \,, \quad u_2 = \frac{1}{\cosh^2(\tau)}\,, \quad\, u_3 =\left.u_1\right|_{\sigma\to -\sigma} \,. \la{ccr}
\eeq
In particular, as an expansion at large $\tau$, we have
\beq
\log u_1 = \sum_{n=0}^{\infty} c_{1,n} e^{-n \tau}\,,  
\qquad \log u_2 = -2 \tau +  \sum_{n=0}^{\infty} c_{2,n} e^{-n \tau} \,,\qquad \log u_3 = \sum_{n=0}^{\infty} c_{3,n} e^{-n \tau}\,. \nn
\eeq
so that \textit{computing the term linear in $\tau$ } is the same (up to a factor of $-2$) as \textit{computing the discontinuity in $u_2$} \cite{Hexagonpaper}.
Finally, it is also useful to have an inverse map relating directly $\tau$, $\sigma$ and $\phi$ to cross-ratios,
\beqa
e^{2\sigma} =\frac{\<1236\>\<1245\>\<3456\>}{\<1256\>\<1346\>\<2345\>} \, , \quad\sinh^2(\tau) =\frac{\<3451\>\<4612\>}{\<3412\>\<4561\>} \, , \quad e^{2\phi} =\frac{\<1236\>\<1456\>\<2345\>}{\<1234\>\<1256\>\<3456\>} \la{inversemap}
\eeqa
\subsection{Details on the Casimir operator(s)} \la{A2}
In this section we give a detailed derivation of the $SL(2)$ Casimir differential operator $\mathbb{C}^{(ijkl)}$ acting on a conformal invariant function $\mathcal{H}^{(ijkl)}$ carrying helicity weight $\eta_i\eta_j\eta_k\eta_l$ (\ref{ratio}). For concreteness, we will follow the example where $(ijkl)=(2256)$ and $\mathbb{C}^{(2256)}$ takes the form   (\ref{diffoperator2}). The generalization ot other components will be clear from the text. 

The tree level ratio function for the component (2256) is
\beq
\cR^{(2256)}[Z_i]   =\frac{\<3561\>}{\<2356\>\<1256\>}
=   -\frac{\text{csch}(\tau )e^{\sigma +\phi }+\coth (\tau )}{
   4\cosh (\sigma ) \cosh (\tau )+4\cosh (\phi )} \la{exampleap}
\eeq
The SL(2) transformation which leaves the null lines associated to the twistors $Z_1$ and $Z_4$ invariant reads 
\beq
g=\left(
\begin{array}{cccc}
 1 & 0 & 0 & 0 \\
 0 & 1 & 0 & 0 \\
 0 & 0 & a & b \\
 0 & 0 & c & \frac{b c+1}{a}
\end{array} 
\right)
\eeq
We then act with this transformation on the bottom twistors $Z_2$ and $Z_3$. We use the notation
\beq\la{gtrans}
\<1256\>_{g} = \det \{Z_1, g Z_2 , Z_5, Z_6 \} 
\eeq
etc. Then we define the action of $SL(2)$ on $\tau$, $\sigma$ and $\phi$ using (\ref{inversemap}). For example
\beq
g\circ \sigma =  \log\sqrt{\frac{\<1236\>_g\<1245\>_g\<3456\>_g}{\<1256\>_g\<1346\>_g\<2345\>_g}}  = \sigma + \log \sqrt{\frac{1+(a-c)(a e^{2\tau}-b)}{1+(a+c)(a e^{2\tau}+b)}}
\eeq
and so on. In particularly, we find that $g\circ \phi=\phi$. Two important quantities are
\beqa
 \mathcal{R}^{(2256)}[Z_1,g Z_2,g Z_3,Z_4,Z_5,Z_6]&=&\frac{\<3561\>_g}{\<2356\>_g\<1256\>_g} \\
 \mathcal{R}^{(2256)}[Z(g\circ\tau,g\circ\sigma,g\circ\phi)]&=&-\frac{\text{csch}(g\circ\tau )e^{g\circ\sigma +\phi }+\coth (g\circ\tau )}{
   4\cosh (g\circ\sigma ) \cosh (g\circ\tau )+4\cosh (\phi )}\nn
\eeqa 
Note that, despite (\ref{exampleap}), these two quantities are not the same! This is the most important point. Their ratio is a nontrivial function of $a,b,c$. Then we consider a general conformal invariant function $\mathcal{H}^{(2256)}$ of the $Z_i$'s carrying helicity weight $(2256)$. For the family of twistors $\{ Z_{2,3}=M(\tau,\sigma,\phi)\cdot \widetilde Z_{2,3},\ Z_a= \widetilde Z_a\text{ for }a\neq 2,3\}$ it can be represented by a function of the three parameters $\tau,\sigma$ and $\phi$, $\mathcal{H}^{(2256)}(\tau,\sigma,\phi)$. Under the SL(2) transformation (\ref{gtrans}) of $Z_2$ and $Z_3$, it transforms as
\beq
g\circ \mathcal{H}^{(2256)}(\tau,\sigma,\phi)=\mathcal{H}^{(2256)}(g\circ\tau,g\circ\sigma,g\circ\phi){ \mathcal{R}^{(2256)}[Z_1,gZ_2,gZ_3,Z_4,Z_5,Z_6]\over \mathcal{R}^{(2256)}[Z(g\circ\tau,g\circ\sigma,g\circ\phi)]}
\eeq
The second factor on the right hand side captures the transformation of the helicity factor. 

Now, let $g_\theta$ be the SL(2) transformation (\ref{gtrans}) with
\beq
 \left(
\begin{array}{cc}
 a & b \\
 c & \frac{b c+1}{a}
\end{array}
\right) = e^{\theta \sigma_i}
\eeq
where $\sigma_i$ are the Pauli matrices. The action of the SL(2) Casimir on $ \mathcal{H}^{(2256)}$ reads
\beqa
&&\mathbb{C}^{(2256)} \circ \mathcal{H}^{(2256)} =\left.\sum_{a=1}^3  \partial_{\theta}^2\,g_\theta\circ  \mathcal{H}^{(2256)}\right|_{\theta=0} \\
&=&\(\[\partial_{\tau}^2+2\coth(2\tau)\partial_{\tau}+{\rm sech}^2(\tau) \partial_{\sigma}^2\]+\[{\rm sech}^2(\tau)(1-2\d_\sigma)-\text{csch}^2(\tau)\]\)\mathcal{H}^{(2256)}\nn\eeqa
For other components, the action of the Casimir is obtained is the same way. Since the transformation of $ \mathcal{R}^{(ijkl)}$ is different for different components, so is the differential operator $\mathbb{C}^{(ijkl)}$. For the $2 3 5 6$ components for example, it turns out that the total helicity weight is invariant under the $SL(2)$ action. As a result, $H^{(2356)}$ transforms as a function of the conformal cross ratios. This is why the differential operator $\mathbb C^{(2356)}$ (\ref{diffoperator}) is exactly the same as the one we found for the hexagon reminder function in \cite{Hexagonpaper}.  

\subsection{Explicit solution to the Ward identities}

Ward identities, studied in \cite{Elvang} and in section \ref{sec3}, allow us to write the NMHV functional in terms of five components alone. This is extremely useful for the OPE program. It allow us to study any component in any channel by choosing appropriately the base of five components. For example, for the choice (\ref{base}) we have 
\begin{eqnarray}\mathcal{R}^{\text{NMHV}}&=&[6,1,2,3,5] \[\frac{\cR^{(2355)} \langle 1235\rangle  \langle 2345\rangle }{\langle
   1234\rangle }+\frac{\cR^{(2356)} \langle 1235\rangle  \langle 2346\rangle }{\langle 1234\rangle
   }\]\nn\\
   &-&[6,2,3,4,5] \[\frac{\cR^{(2355)} \langle 1235\rangle  \langle 2345\rangle }{\langle 1234\rangle
   }+\frac{\cR^{(2356)} \langle 1236\rangle  \langle 2345\rangle }{\langle 1234\rangle
   }\]\nn\\
   &+&[6,1,2,4,5] \[\cR^{(2356)}\(\frac{\langle 2346\rangle  \langle 2351\rangle }{\langle 1234\rangle
   }+\frac{\langle 5612\rangle  \langle 5643\rangle }{\langle 1456\rangle }\right)\right.\nn\\
   &&\left.\qquad\qquad\qquad-\,\frac{\cR^{(2256)} \langle 1256\rangle  \langle 2456\rangle }{\langle
   1456\rangle }-\frac{\cR^{(2355)} \langle 1235\rangle  \langle 2345\rangle }{\langle 1234\rangle
   }\]\la{ward}
\\
   &-&[6,1,2,3,4]
   \[\cR^{(2356)} \left(\frac{\langle 1236\rangle  \langle 2345\rangle }{\langle 1234\rangle
   }+\frac{\langle 1235\rangle  \langle 2346\rangle }{\langle 1234\rangle }\right)\right.\nn\\
   &&\left.\qquad\qquad\qquad+\,\frac{\cR^{(2355)} \langle 1235\rangle  \langle 2345\rangle }{\langle 1234\rangle
   }+\frac{\cR^{(2366)}
   \langle 1236\rangle  \langle 2346\rangle }{\langle 1234\rangle }\]\nn\\
   &+&[6,1,3,4,5]
   \[\cR^{(2356)} \left(\frac{\langle 5612\rangle  \langle 5643\rangle }{\langle 1456\rangle
   }-\frac{\langle 2345\rangle  \langle 2361\rangle }{\langle 1234\rangle }\right)\right.\nn\\
   &&\left.\qquad\qquad\qquad+\,\frac{\cR^{(2355)} \langle 1235\rangle  \langle 2345\rangle }{\langle 1234\rangle
   }-\frac{\cR^{(3356)}
   \langle 1356\rangle  \langle 3456\rangle }{\langle 1456\rangle }\]\nn
\end{eqnarray}

\end{document}